\documentclass[12pt]{article}
\usepackage{amsmath}
\usepackage{amsfonts}        
\usepackage{amssymb}
\usepackage{amsbsy}
\usepackage{float}

\usepackage{mathrsfs}
\usepackage{epsfig}      
\usepackage{xcolor}       

\usepackage{bm}          

\usepackage[T1]{fontenc} 

\usepackage{graphicx}    

\newlength{\TZ}
\DeclareFontFamily{OT1}{pzc}{}
\DeclareFontShape{OT1}{pzc}{m}{it}{<-> s * [1.200] pzcmi7t}{}
\DeclareMathAlphabet{\mathpzc}{OT1}{pzc}{m}{it}
\newcommand{\BEQ}{\begin{equation}}     
\newcommand{\BEA}{\begin{eqnarray}}
\newcommand{\BD}{\begin{displaymath}}
\newcommand{\EEQ}{\end{equation}}       
\newcommand{\EEA}{\end{eqnarray}}
\newcommand{\ED}{\end{displaymath}}

\newcommand{\D}{{\rm d}}                
\newcommand{\II}{{\rm i}}               
\newcommand{\wit}[1]{\widetilde{#1}}    
\newcommand{\wht}[1]{\widehat{#1}}      
\newcommand{\ket}[1]{\left|#1\right\rangle}  

\newcommand{\vek}[1]{\bm{#1}}           

\newcommand{\fns}{\footnotesize}        
\newcommand{\scs}{\scriptsize}          
\newcommand{\hs}[1]{\hspace{-#1truecm}}   

\newcommand{\appsektion}[1]{\setcounter{equation}{0}\setcounter{subsection}{0}
\section*{Appendix. #1}
\renewcommand{\theequation}{A.\arabic{equation}}
              \renewcommand{\thesection}{A}
              \renewcommand{\thefigure}{A\arabic{figure}}\setcounter{figure}{0}
              \renewcommand{\thetable}{A\arabic{table}}\setcounter{table}{0} }

\catcode`\@=11
\def\numberbysection{\@addtoreset{equation}{section}
        \def\theequation{\thesection.\arabic{equation}}}
\numberbysection                        

\definecolor{gruen}{rgb}{0,0.625,0}       
\definecolor{rot}{rgb}{0.75,0,0}          
\definecolor{blau}{rgb}{0,0,0.75}         
\definecolor{casta}{rgb}{0.45,0.20,0}     
\definecolor{gelb}{rgb}{0.825,0.725,0.0}  

\newcommand{\blau}[1]{\textcolor{blau}{#1}}       
\newcommand{\rot}[1]{\textcolor{rot}{#1}}         
\newcommand{\gruen}[1]{\textcolor{gruen}{#1}}     
\newcommand{\laran}[1]{\textcolor{orange}{#1}}    
\newcommand{\mage}[1]{\textcolor{magenta}{#1}}    

\begin{document}

\begin{titlepage}

\vskip 1.5 cm
\begin{center}
{\LARGE \bf Short-time dynamics in phase-ordering kinetics}
\end{center}

\vskip 2.0 cm
\centerline{{\bf Le\"{\i}la Moueddene}$^{a,b\,}$\footnote{Address after 1$^{\rm st}$ of November: 
Physikalisches Institut, Albert-Ludwigs-Universit\"at Freiburg, Herrmann-Herder Stra{\ss}e 3, D-79104 Freiburg i. Breisgau, Germany \\}
and {\bf Malte Henkel}$^{a,c}$}
\vskip 0.5 cm
\centerline{$^a$Laboratoire de Physique et Chimie Th\'eoriques (CNRS UMR 7019),}
\centerline{Universit\'e de Lorraine Nancy, B.P. 70239, F -- 54506 Vand{\oe}uvre l\`es Nancy Cedex, France}

\vspace{0.5cm}
\centerline{$^b$Centre for Fluid and Complex Systems, Coventry University,}
\centerline{Coventry CV1 5FB, United Kingdom}
\vspace{0.5cm}

\centerline{$^c$Centro de F\'{i}sica Te\'{o}rica e Computacional, Universidade de Lisboa,}
\centerline{Campo Grande, P -- 1749-016 Lisboa, Portugal}
\vspace{0.5cm}

\begin{abstract}
Short-time dynamics in the $2D$ Blume-Capel model, with a non-conserved or\-der-pa\-ra\-me\-ter and short-ranged interactions, is analysed. 
For non-equilibrium dynamics, both at a critical point in the $2D$ Ising universality class and at the tricritical point,
we reproduce the values $\Theta=0.190({5})$ and $\Theta=-0.542({5})$, respectively, of the critical initial slip exponent. 
These agree with more early estimates and with the Janssen-Schaub-Schmittmann scaling relation. 
In phase-ordering kinetics, after a quench into the ordered phase, we establish the validity of short-time dynamics. 
In the $2D$ Ising universality class, we find $\Theta=0.39({1})$ in agreement with the scaling relation $\lambda=d-2\Theta$. 
\end{abstract}
\end{titlepage}

\setcounter{footnote}{0}

\section{Introduction} \label{sec:1}

Elucidating from short-time data the behaviour of complex systems at late times is a long-standing objective in 
many branches of natural sciences \cite{Stef25,ElHage17,Mosca17,Giam16,Chen20,Mani21,Bait22,Boro24,Duva24,Luo25}.  
We shall be focussed on far-from-equilibrium systems, notably those which undergo physical ageing \cite{Stru78,Vinc24}. 
Implicitly, we shall always work in the context of `model-A-type dynamics' without any macroscopic conservation law \cite{Taeu14}. 
In practise, ageing is brought about by initially preparing a (classical) many-body system in a fully disordered high-temperature state 
before quenching it instantaneously either onto a critical point $T=T_c$ \cite{Zhen98,Godr02,Cugl03,Cala05,Puri09,Henk10,Alba11,Taeu14} 
or else into the ordered phase with temperature $T<T_c$ \cite{Bray94a,Cugl03,Maze06,Puri09,Henk10,Cugl15}.
When $T=T_c$, one speaks of {\em non-equilibrium critical dynamics} \cite{Godr02}
whereas for $T<T_c$, one is dealing with {\em phase-ordering kinetics} \cite{Bray94a}. 
Phenomenologically, ageing systems obey the three defining properties \cite{Henk10}: 
(A) slow relaxational dynamics, (B) absence of time-translation-invariance and (C) dynamical scaling.
All three properties are required to specify the physical phenomenon we have in mind. 
Microscopically, and after a quench onto $T=T_c$, the dynamics is characterised by critical-point fluctuations \cite{Cala05,Taeu14},
whereas after a quench into $T<T_c$ there is a competition of at least two distinct, but equivalent, 
equilibrium states and the dynamics is driven by the interface tension between them \cite{Bray94a}.
In either case, the system becomes spatially non-homogeneous and decomposes microscopically into clusters
of a time-dependent and growing linear size $\ell=\ell(t)$.
We shall concentrate on situations where this growth is algebraic at late times, viz. $\ell(t)\sim t^{1/\mathpzc{z}}$, 
which defines the dynamical exponent $\mathpzc{z}$.
Its value will be different for $T=T_c$ or $T<T_c$. 
It is often adequate to use a time-space-dependent order-parameter $\phi(t,\vek{r})$, 
in the continuum limit.\footnote{The terms `short-time' and `long-time' are meant to refer to the continuum
limit. We shall always consider times large with respect to microscopic time-scales $\tau_{\rm mic}$ 
needed before arriving at the dynamical scaling regimes we are interested in.} 
For convenience, we shall speak in terms of magnetic systems. 
Then, for pure magnets, the order-parameter is the the coarse-grained local magnetisation.

In a ground-breaking advance, it was recognised by Janssen, Schaub and Schmittmann ({\sc jss}) \cite{Jans89} 
in non-equilibrium critical dynamics, after a quench onto $T=T_c$, that
for an initial state which has a small non-vanishing magnetisation $m_0$, but whose spatial correlations are very short-ranged,  
there exists a novel short-time scaling regime. They showed that the average magnetisation evolves as 
\BEQ \label{eq:JanssenM}
M(t) = m_0\, t^{\Theta} \mathscr{F}_M\left( m_0\, t^{\Theta +\beta/(\nu \mathpzc{z})}\right) \;\; , \;\;
\mathscr{F}_M(u) \simeq \left\{ \begin{array}{ll} \mathscr{F}_0               & \mbox{\rm ~~;~ if $u\ll 1$} \\
                                                  \mathscr{F}_{\infty} u^{-1} & \mbox{\rm ~~;~ if $u\gg 1$} 
                              \end{array} \right.
\EEQ
where $\mathscr{F}_{0,\infty}$ are constants and $\Theta$ is the {\em initial slip exponent}\footnote{In the literature, 
the initial slip exponent is denoted differently, viz. $\theta,\theta',x_0,\ldots$, by different schools. 
We strive for clarity and use a different notation in order to avoid any confusions.}  
which is independent of any equilibrium critical exponents such as $\beta,\nu,\eta,\mathpzc{z}$, at least for model-A dynamics onto which this paper will exclusively focus. 
The scaling function $\mathscr{F}_M(u)$ interpolates between the initial power-law $M(t)\sim t^{\Theta}$ 
for short times and the distinct power-law decay $M(t)\sim t^{-\beta/\nu\mathpzc{z}}$ for late times, which is independent of $m_0$. 
 \begin{figure}[tb]
     \centering
     \includegraphics[width=01\linewidth]{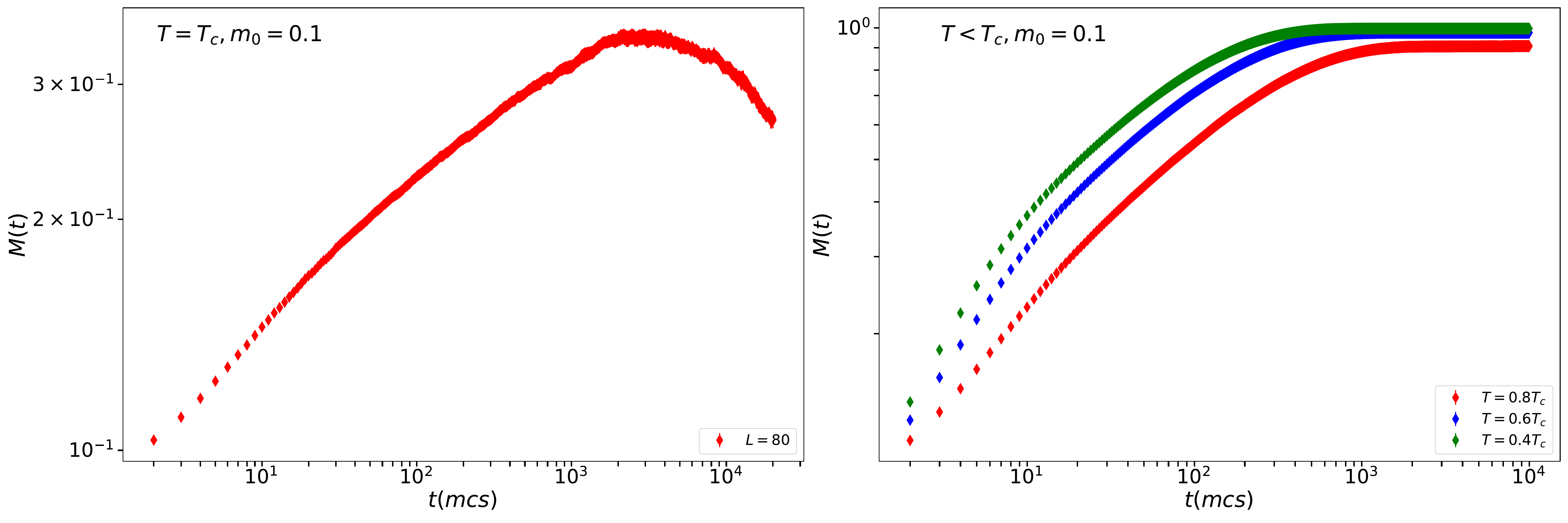}
     \caption[fig2]{Magnetisation in the $2D$ Blume-Capel model, for a periodic square lattice $L\times L$ with $L=80$ and the initial magnetisation $m_0=0.1$. 
     \underline{Left panel:} Initial growth $M(t)\sim t^{\Theta}$ at the critical point $P_c$ (at $T=T_c$, $\Delta_c=0$) 
     and the cross-over towards the equilibrium-controlled decay. The initial slip exponent is $\Theta\approx 0.19$. 
     \underline{Right panel:} Initial growth below the critical point, for $T=[0.4T_c, 0.6T_c, 0.8T_c]$ 
     from top to bottom, and the cross-over towards saturation. The initial slip exponent is $\Theta\approx 0.39$.}
     \label{fig:aimantation}
 \end{figure}
In figure~\ref{fig:aimantation}, we illustrate in the left panel the initial growth of the magnetisation 
$M(t)$ for a model in the $2D$ Ising universality class at $T=T_c$ 
(the data come from the $2D$ Blume-Capel model, to be defined in section~\ref{sec:2}). For larger times, there is a
cross-over visible, towards the late-time decay. The value of the universal exponent $\Theta$ can be used, 
as the equilibrium critical exponents, to characterise the dynamical universality class.
Since this behaviour is not affected by the system not having equilibrated, 
and also is quite insensitive to finite-size effects, it has become a very widely used technique, see \cite{Zhen98,Alba11} 
for classical reviews.  The range of applications
has further increased ever since; an incomplete list on recent developments 
might include studies on critical and tricritical points \cite{Silva13,Silva13b,Liu16,Wanz18,Vodr22,Silva22a,Silva23,Ihss25}, 
non-equilibrium systems without detailed balance \cite{Anjo20,Nasc21,Silva25a}, simple liquids and their spinodal lines \cite{Losc16}, 
models of catalytic reactions \cite{Fern25}, active matter \cite{Puzz22} and imaginary-time quantum dynamics \cite{Shu17,Shu22,Wild23,Faeh25} of closed systems. 
The method of short-time dynamics nowadays belongs to the standard toolbox
of the statistical physicist for the study of non-equilibrium critical dynamics.

Analogously, for higher moments $M^{(n)}(t)$, with integer $n\geq 2$, and for spatially infinite systems, one has a scaling form 
$M^{(n)}(t) = m_0^n\, t^{n\Theta} \mathscr{F}^{(n)}_M\bigl( m_0\, t^{\Theta +\beta/(\nu \mathpzc{z})}\bigr)$ with $\mathscr{F}^{(n)}_M(0)=\mbox{\rm cste.}$ 
and $\mathscr{F}^{(n)}_M(u)\sim u^{-n}$ for $u\gg 1$, see \cite{Alba11,Silva13b}.\footnote{On  lattices of finite linear size $L$, one usually quotes 
$M^{(2)}(t;L)\sim L^{-d} t^{(d-2\beta/\nu)/\mathpzc{z}} = L^{-d} t^{(2-\eta)/\mathpzc{z}}$ \cite{Okan97,Okan97b}.}  
One may also study global correlation functions, for example the global correlator of a time-space magnetisation $M(t,\vek{r})$ 
with the initial state $M(0,\vek{0})$, even for an initially
unmagnetised state with $m_0=0$. Then the scaling of the global correlator 
$C(t) = \sum_{\vek{r}} \bigl\langle M(t,\vek{r})M(0,\vek{0})\bigr\rangle \sim t^{\Theta}$ \cite{Tome98,Tome03} 
for short times is governed by the initial slip exponent as well. The initial slip exponent obeys the celebrate {\sc jss} scaling relation \cite{Jans89}
\BEQ \label{eq:jss} 
\lambda = d - \mathpzc{z} \Theta
\EEQ
where the {\em auto-correlation exponent} $\lambda$ \cite{Huse89} describes the decay of the local auto-correlator 
$A(t)=\bigl\langle M(t,\vek{0})M(0,\vek{0})\bigr\rangle\sim t^{-\lambda/\mathpzc{z}}$ for late times
(and always an unmagnetised initial state). Eq.~(\ref{eq:jss}) provides therefore a link between the short-time exponent $\Theta$ and the long-time exponent $\lambda$. 
This constitutes an example of the looked-after relationships between short-time and long-time dynamics mentioned above. 
An exactly solvable example is the $1D$ Time-Dependent Ginzburg-Landau equation at $T=0$ \cite{Bray95}.
While the original derivation \cite{Jans89} of the scaling form (\ref{eq:JanssenM}) and the scaling relation (\ref{eq:jss}) required elaborate field-theoretic considerations 
based on the short-time operator product expansion, it was realised recently that 
these results also follow from the twin hypothesis' of dynamical scaling and time-translation-invariance, 
provided both symmetries are used in a new time-dependent non-equilibrium representation \cite{Henk25,Henk25c}. 
This construction implies that a scaling operator $\phi$ is not only characterised by an
equilibrium scaling dimension $\delta$, but also by a second, independent scaling dimension $\xi$,  related to this change of representation.

{\em Short-time critical dynamics does require the existence of dynamical scaling} to come into being. For phase-ordering kinetics, 
obtained after a quench into the ordered phase at $T<T_c$ with at least two distinct, but equivalent equilibrium states, dynamical scaling holds as well \cite{Bray90,Bray94a}. 
For $T<T_c$, being far away from the critical point, one might believe that the system should saturate rapidly whenever $m_0\ne 0$. But this is not so, as shown
in the right panel of figure~\ref{fig:aimantation} which displays the evolution of $M(t)$ for several temperatures $T$ far below the critical temperature $T_c>0$. 
Surprisingly, we recognise the existence of a short-time scaling regime where $M(t)\sim t^{\Theta}$. However, the initial slip exponent 
$\Theta$ has a value distinct from the one at criticality. As we shall show, for an initial state with magnetisation $m_0$ but without
spatial correlations, the following scaling form can be derived for $T<T_c$ 
\BEQ \label{eq:MbasT}
M(t) = m_0\, t^{\Theta} \mathscr{F}_M\left( m_0^{y_0}\, t \right) \;\; , \;\; 
\mathscr{F}_M(u) \sim \left\{ \begin{array}{ll} 1           & \mbox{\rm ~~;~ if $u\ll 1$} \\
                                                u^{-\Theta} & \mbox{\rm ~~;~ if $u\gg 1$} 
                              \end{array} \right.
\EEQ
which now replaces (\ref{eq:JanssenM}) and where $y_0$ is related to the scaling dimension of $m_0$. For small times, 
this reproduces the algebraic growth with the time $t$, whereas for large times, the magnetisation $M(t)\stackrel{t\to\infty}{\to} M_{\infty}(m_0)\sim m_0^{\mu_0}$ 
saturates, with $\mu_0=1-\Theta y_0<1$. These are the qualitative features seen in the right panel of figure~\ref{fig:aimantation}. 
Clearly, the shapes of the scaling functions are distinct for $T<T_c$ and $T=T_c$. 
Remarkably, the scaling relation (\ref{eq:jss}) relating the short-time exponent $\Theta$ and the long-time exponent $\lambda$ continues to hold true for $T<T_c$ \cite{Henk25c}. 
Of course, the exponents $\Theta$, $\lambda$ and $\mathpzc{z}$ take values different from the ones at the critical point.\footnote{In the phase-ordering $2D$ Ising model, 
the non-rigorous Fisher-Huse inequality $\lambda \leq \frac{5}{4}$ \cite{Fish88a} would via (\ref{eq:jss}) imply the bound $\Theta\geq \frac{3}{8}$. 
The Fisher-Huse conjecture \cite{Fish88a} states $\lambda=\frac{5}{4}$, hence $\Theta=\frac{3}{8}$.} 
For example, in the case of model-A-type dynamics without a conservation law for the order-parameter $\phi(t,\vek{r})$, 
and with short-ranged interactions (implicitly assumed throughout), it is well-known that
$\mathpzc{z}=2$ \cite{Bray94b}.

This work is organised as follows. 
Section~\ref{sec:2} defines the Blume-Capel model and gives its main properties. We also define the observables we are going to study. 
Section~\ref{sec:3} analyses short-time dynamics for critical quenches, both for the $2D$ Ising universality class and the tricritical point of the $2D$ Blume-Capel model. 
The exponent $\Theta$ is found and we test and confirm the {\sc jss} scaling relation (\ref{eq:jss}). 
Section~\ref{sec:4} studies, for the first time, short-time dynamics in phase-ordering kinetics after a quench into $T<T_c$. We conclude in section~\ref{sec:5}. 
The appendix derives the short-time dynamics (\ref{eq:MbasT}) of the local magnetisation $M(t)$ in phase-ordering kinetics, along with the scaling relation (\ref{eq:jss}).

\section{The Blume-Capel model} \label{sec:2}

\begin{figure}[tb]
     \centering
     \includegraphics[width=0.5\linewidth]{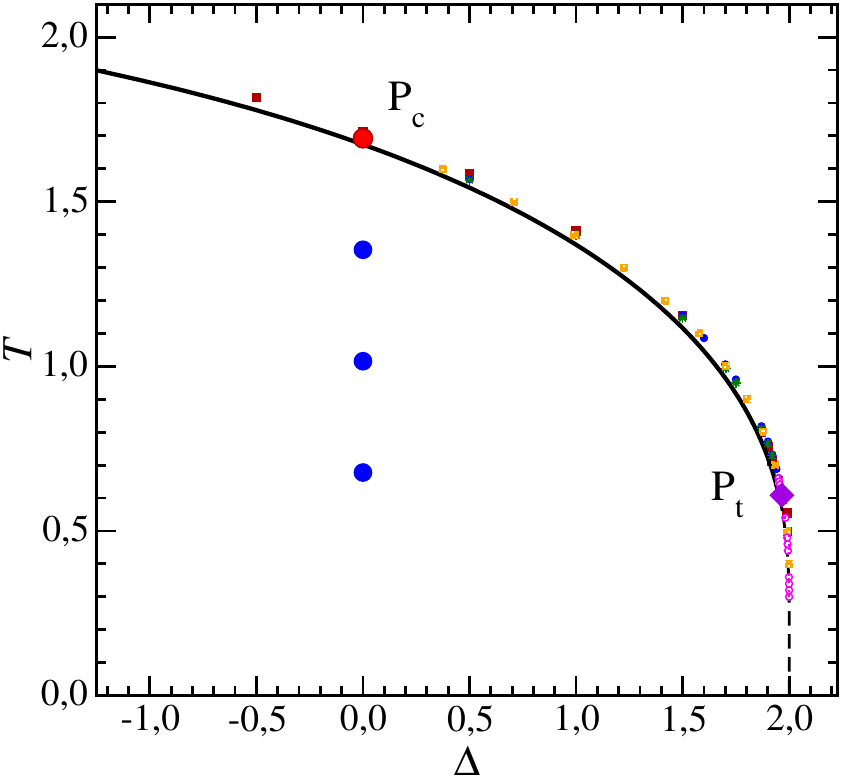}
     \caption[fig1]{Phase diagram of the $2D$ Blume-Capel model. 
     The prediction (\ref{eq:BC-lignecrit}) is the full line for the second-order transitions and the dashed line for the first-order transitions, 
     separated by the tricritical point $P_{\rm t}$ at $(T_{\rm t}=0.608, \Delta_{\rm t}=1.966)$ (violet diamond). 
     The Ising critical point $P_c$ is at $(T=T_c=1.6929, \Delta=0)$ (red dot).  
     The three blue dots at $\Delta=0$ give the locations for the low-temperature measurements. The small symbols on the critical line indicate numerical estimates, obtained from
     Wang-Landau Monte Carlo simulation (red square {\tiny\rot{$\blacksquare$}} \cite{Silva}, open magenta circle {\fns\mage{$\circ$}} \cite{Kwak15}), 
     microcanonical annealing (blue circle {\fns\blau{$\bullet$}} \cite{Mozo24}), 
     high-temperature series (green star {\fns\gruen{$\ast$}} \cite{Bute18}) and hybrid Monte Carlo (orange square {\tiny\laran{$\blacksquare$}} \cite{Zier17}).  
     }
     \label{fig:2DBCphasediag}
 \end{figure}

The $2D$ Blume-Capel model \cite{Lawr84} is defined, on a square $L\times L$ lattice $\Lambda\subset \mathbb{Z}^2$, by the hamiltonian  
\begin{equation} \label{eq:BCHam} 
\mathcal{H}=-J\sum_{\langle i,j\rangle}\sigma_i \sigma_j + \Delta \sum_i \sigma_i^2 
\end{equation}
where each spin variable $\sigma_i$ on site $i\in \Lambda$ can take the values $\sigma_i \in \{ -1, 0 , +1\}$. 
We shall use nearest-neighbour interactions and periodic boundary conditions.  
The first term in (\ref{eq:BCHam}) represents the ferromagnetic exchange interaction of strength $J>0$ (we scale to $J=1$), 
and the second term, governed by the crystal-field parameter $\Delta$, controls the concentration of vacancies corresponding to sites with $\sigma_i=0$. 
The $2D$ Blume–Capel model has been widely investigated in equilibrium, primarily through numerical 
simulations~\cite{Silva,Qian05,Mala10,Kwak15,Zier17,Bute18,Luqu19,Vata20,Mend24,Mozo24,Mata25,Mata25a,Mace24,Shi24,Moue24,Moue25,Henr25,Liu25}. 
There exists an analytical prediction for the critical line derived from the hamiltonian (\ref{eq:BCHam}) \cite{Clus08}
\BEQ \label{eq:BC-lignecrit}
\frac{\Delta}{J} = \frac{T}{J} \ln\left( 2 \sinh\frac{2J}{T} - 2 \right)
\EEQ
In figure~\ref{fig:2DBCphasediag} this prediction is compared to various recent numerical results \cite{Silva,Kwak15,Zier17,Bute18,Mozo24}.
For $\Delta=0$, there is a second-order phase-transition at a critical point at the critical temperature $T_c\simeq 1.6929$ (red dot), 
which is in the universality class of the $2D$ Ising model. We shall call this point $P_c$. 
We shall also investigate the behaviour at smaller values of $T<T_c$, which are also indicated by the blue points in figure~\ref{fig:2DBCphasediag}. 
When $\Delta$ is increased, the concentration of vacancies 
(sites where $\sigma_i=0$) also increases and there is a line of $2D$ Ising transitions which ends in a tri-critical point (purple rhombus) \cite{Lawr84} 
at $T_{\rm t}= 0.60857756(4), \Delta_{\rm t}\simeq 1.9658149(2)$ \cite{Qian05,Bute18}. This point is called $P_{\rm t}$. 
Beyond, there is a line of first-order transitions (dashed line).  The analytical prediction (\ref{eq:BC-lignecrit}) agrees with the numerical results within $1\%$.

The dynamics of the model, with a non-conserved order-parameter, is generated via a standard Monte Carlo dynamics of 
Metropolis type.\footnote{The universality of non-equilibrium dynamics with respect to different types of local model-A-dynamics is well-known \cite{Okan97,Okan97b}.}  
The spins become time-dependent $\sigma_i=\sigma_i(t)$. At criticality $T=T_c$ and below (at $T=0.4T_c, 0.6T_c, 0.8T_c$, see the blue points in figure~\ref{fig:2DBCphasediag}), 
simulations were performed using six independent runs, each consisting of 1000 samples, for both $m_0 = 0$ and $m_0 \neq 0$, 
with system sizes $L = 80$ and $L = 160$. Here, independent runs use different random number sequences for error estimation, while samples are independent measurements from independent initial configurations. At the tricritical point, twelve independent runs of 4000 samples each were carried out 
to improve statistical accuracy. The statistical error bars correspond to the standard deviation among the independent runs.

In contrast, the non-equilibrium dynamics, notably at the tricritical point, 
has been explored thoroughly by da Silva {\it et al.} \cite{Silva02,Silva13,Silva14}, see also \cite{Liu23}, 
and we refer to \cite{Alba11} for more old studies. 

We shall consider the following observables, where ${\cal N}=|\Lambda|=L^2$ is the total number of sites of the lattice $\Lambda$ 
\begin{subequations} \label{eq:obs} \begin{align} 
\mbox{\rm magnetisation~~~~~~} 
       M(t) &= \frac{1}{\cal N}\sum_{i\in\Lambda} \bigl\langle \sigma_{i}(t) \bigr\rangle = \bigl\langle \phi(t,\vek{0})\bigr\rangle \\
\mbox{\rm squared magnetisation~~~} 
       M^{(2)}(t) &= \left\langle \left( \frac{1}{\cal N}\sum_{i\in\Lambda} {\sigma_i(t)} \right)^2\, \right\rangle = \left\langle \phi(t,\vek{0})^2\right\rangle \\
\mbox{\rm global correlator~~~~~~}  
       C(t) &= \frac{1}{{\cal N}^2}\sum_{i,j\in\Lambda} 
                \biggl\langle \bigl( \sigma_i(t) - \langle \sigma_i(t)\rangle\bigr)\bigl( \sigma_j(0) - \langle \sigma_j(0)\rangle\bigr) \biggr\rangle
                \label{eq:obs-C} \\
            &= \frac{1}{\cal N}\int_{\mathbb{R}^{d}} \!\D \vek{r}\: 
                \biggl\langle \bigl( \phi(t,\vek{r}) - \langle \phi(t,\vek{r})\rangle \bigr) \bigl( \phi(0,\vek{0}) - \langle \phi(0,\vek{0})\rangle \bigr)\biggr\rangle 
                \nonumber \\
\mbox{\rm autocorrelator~~~~~~} 
       A(t) &= \frac{1}{\cal N}\sum_{i\in\Lambda} 
                \left. \bigl\langle \sigma_i(t) \sigma_i(0) \bigr\rangle 
                \right|_{m_0=0} 
            = \left.\bigl\langle \phi(t,\vek{0})\phi(0,\vek{0})\bigr\rangle \right|_{m_0=0}
            \label{eq:obs-A}
\end{align}\end{subequations}
Herein, we give both the explicit construction from the discrete spins $\sigma_i(t)$ 
and also their interpretation in terms of the continuum order-parameter $\phi(t,\vek{r})$. 
The respective equalities are expected to hold in the continuum limit. 
The magnetisation $M(t)$ is only non-vanishing for $m_0\neq 0$, as follows from (\ref{eq:JanssenM}). 
For $M^{(2)}(t)$ theoretical predictions shall depend on whether $m_0$ vanishes or not. 
For the global correlator $C(t)$ admitting $m_0\neq 0$ can improve the quality of the lattice data, as we shall discuss below. 
For $m_0=0$, the connected global correlator in (\ref{eq:obs-C}) reduces to the standard one 
$C(t)={\cal N}^{-2}\sum_{i,j\in\Lambda} \bigl\langle \sigma_i(t)\sigma_j(0)\bigr\rangle$. 
The auto-correlator $A(t)$ is always evaluated for $m_0=0$.

\section{Quenches onto $T=T_c$} \label{sec:3}

We shall begin by considering quenches onto the critical line of second-order phase transitions, see figure~\ref{fig:2DBCphasediag}, 
which are in the $2D$ Ising universality class with the exception of the tricritical point $P_t$. 
Our consideration of the short-time dynamics merely aims to check
the consistency with the results already available in the literature \cite{Jans89,Gras95,Okan97,Okan97b,Mend98,Tome98,Zhen98,Silva02,Tome03,Silva09,Alba11,Silva13,Taeu14,Silva23} 
and in this way confirm our numerical algorithms. 
We focus therefore on the Ising critical point $P_c$ and then on the Ising tricritical point $P_{\rm t}$.

We shall be concerned with the behaviour in sufficiently large lattices such that finite-size effects will turn out to be unimportant. 
{}From the definition (\ref{eq:obs}), and since we are mainly interested in the short-time regime, 
we expect the critical behaviour \cite{Jans89,Okan97b,Tome98,Tome03,Silva23}
\BEQ \label{eq:asympt}
M(t) \sim t^{\Theta} \;\; , \;\;
M^{(2)}(t) \sim \left\{ \begin{array}{ll} t^{2\Theta}              & \mbox{\rm ~~;~ $m_0\neq 0$} \\ 
                                          t^{(2-\eta)/\mathpzc{z}} & \mbox{\rm ~~;~ $m_0=0$}  
                        \end{array} \right. \;\; , \;\;
C(t) \sim t^{\Theta} \;\; , \;\;
A(t) \sim t^{-\lambda/\mathpzc{z}} 
\EEQ 
Herein, $M(t), M^{(2)}(t), C(t)$ are studied in the short-time regime, while the auto-correlator $A(t)$ is computed in the long-range regime. 
The magnetisation $M(t)$ is always studied for $m_0\neq 0$, for
the squared magnetisation $M^{(2)}(t)$ the expected behaviour depends on $m_0$ as indicated, for $C(t)$ the dependence on $m_0$ 
will be discussed below and for $A(t)$ we always have $m_0=0$. We shall check  the scaling relation (\ref{eq:jss}). 
All results (\ref{eq:asympt}) also follow from the combined dilatation- and time-translation-invariance in their new time-dependent non-equilibrium representation \cite{Henk25c}. 

\subsection{The critical case}

\begin{figure}[tb]
  \hspace{-1.45cm} 
     \includegraphics[width=1.18\linewidth]{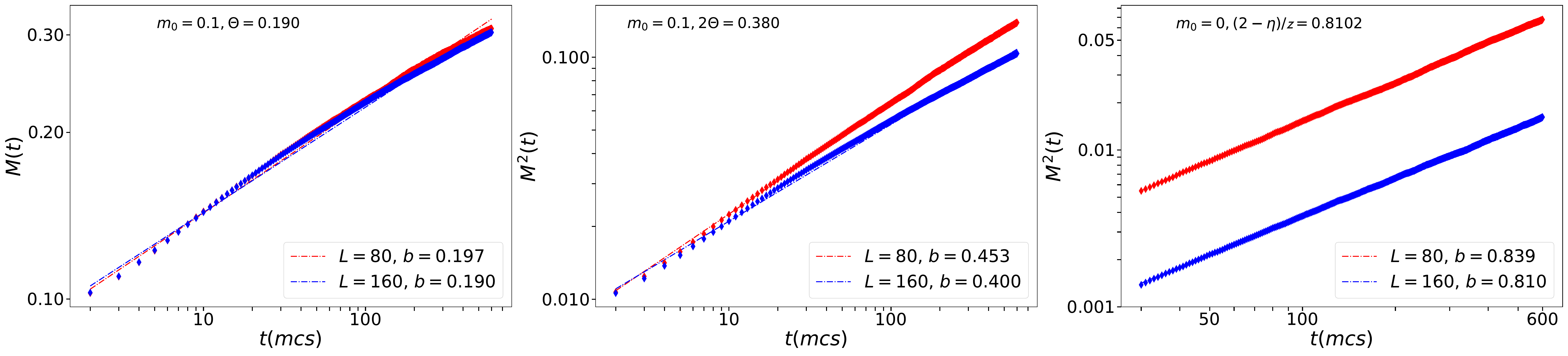}
     \caption[fig3]{Quench onto the point $P_c$, for $L=80$ (red upper curve) and $L=160$ (blue lower curve) 
     and initial magnetisation $m_0$. \underline{Left panel:} Magnetisation $M(t)$ for $m_0=0.1$. Fit range: $t=2-600$. 
     \underline{Centre panel:} Squared magnetisation $M^{(2)}(t)$ for $m_0=0.1$. Fit range: $t=2-600$. 
     \underline{Right panel:} Squared magnetisation $M^{(2)}(t)$ for $m_0=0$. Fit range: $t=30-600$.
     }
     \label{fig:IsingC_aimantations}
 \end{figure}

We first consider a quench of the $2D$ Blume-Capel model onto the Ising critical point $P_c$, the red point in figure~\ref{fig:2DBCphasediag}. 
The magnetisation $M(t)$ and the squared magnetisation $M^{(2)}(t)$ are shown in figure~\ref{fig:IsingC_aimantations}. 
In the left panel, the short-time behaviour of the magnetisation $M(t)$ is shown for two distinct lattice sizes, $L=80$ and $L=160$ respectively, and for $m_0=0.1$. 
Since the two curves are practically super-imposed, we can conclude that the lattices are already so large that finite-size effects are no longer notable. 
A linear fit is made on the logarithmic form $\ln M(t) = a + b \ln t$ where the coefficient $b$ will give an estimate for $\Theta$. This will be done similarly throughout. 
The range of $t$-values used in the respective fits will be indicated in the captions of the respective figures and resulting values of $b$ in the legends.
{}From the algebraic growth (\ref{eq:asympt}),
we read off from the more large value of $L$ the initial slip exponent $\Theta=0.190(5)$ in excellent agreement
with the expected value from the literature \cite{Gras95,Okan97,Okan97b,Mend98,Zhen98,Alba11}, 
see also table~\ref{tab:1}.\footnote{The reviews \cite{Zhen98,Alba11}, and many of their sources, denote the ratio $\lambda/\mathpzc{z}$ by $\lambda$.} 
Similarly, in the centre panel, the short-time evolution of the squared magnetisation $M^{(2)}(t)$ is shown
for the same lattices and the same initial magnetisation. 
For small times, the two curves coincide, but in the smaller lattice we begin to observe the onset of a more rapid growth towards saturation, 
as mediated by finite-size effects. Here only the data for the larger lattice with $L=160$ can be analysed quantitatively and yield 
$2\Theta=0.40(4)$.  

In the right panel of figure~\ref{fig:IsingC_aimantations}, we display once more
the squared magnetisation $M^{(2)}(t)$, but now for an unmagnetised initial state with
$m_0=0$, and for the same lattice size as before. In this case, the scaling (\ref{eq:asympt}) rather predicts the short-time behaviour  
$M^{(2)}(t) \sim t^{(2-\eta)/\mathpzc{z}}$
with the equilibrium critical exponent $\eta$. On the other hand, the scaling amplitude is clearly size-dependent, with a relative factor $L^{-d}$ \cite{Okan97}. 
For the $2D$ Ising universality class, it
is well-known that $\eta=\frac{2\beta}{\nu}=\frac{1}{4}$, see table~\ref{tab:1}, which implies that $\frac{2-\eta}{\mathpzc{z}}\simeq 0.8102$. 
The data for both $L=80$ and $L=160$ show the anticipated initial algebraic increase with $t$, 
and, since the lines are roughly parallel, with sensibly the same numerical value of the exponent. 
However, the data for $L=80$ appear to be slightly tainted by incipient finite-size effects while this does not yet seem to be the case for $L=160$. 
In the latter case, a fit yields $\frac{2-\eta}{\mathpzc{z}}=0.810(4)$ in excellent agreement with the expectation. 
This is the confirmation of both expressions in (\ref{eq:asympt}) for $M^{(2)}(t)$. 

   \begin{figure}[h]
  \hspace{-1.45cm} 
  \includegraphics[width=1.18\textwidth]{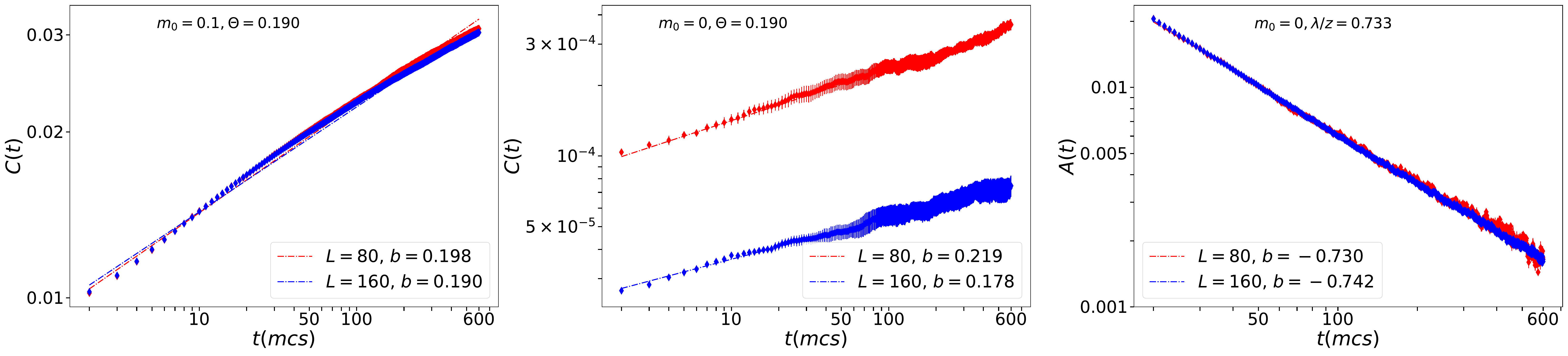}
     \caption[fig4]{Quench onto the point $P_c$, for $L=80$ (red lower curve) and $L=160$ (blue upper curve) and initial magnetisation $m_0$. 
     \underline{Left panel:} Global correlator $C(t)$ for $m_0=0.1$.  
     Fit range: $t=2-600$. 
     \underline{Centre panel:}  Global correlator $C(t)$ for $m_0=0$. 
     Fit range: $t=2-600$. 
     \underline{Right panel:} auto-correlator $A(t)$ for $m_0=0$. Fit range: $t=20-600$. 
     }
     \label{fig:IsingC_correlateur}
 \end{figure}

Next, we show in figure~\ref{fig:IsingC_correlateur} data on global and local correlators.
In the left panel we present data for the global correlator $C(t)$, for the initial magnetisation $m_0=0.1$ 
and the lattice sizes $L=80$ and $L=160$. We observe the expected
algebraic increase (\ref{eq:asympt}) with the time $t$ and the roughly parallel lines indicate similar values of the exponent. 
However, for $L=80$ there is a slight curvature in the data which appear to cross over towards saturation. 
These signals are at least much more weak for $L=160$. 
A fit to the latter data produces $\Theta=0.190(5)$ and is clearly 
in excellent agreement with our above results from the magnetisations and the literature \cite{Okan97,Mend98,Zhen98,Henk10,Alba11}, see also table~\ref{tab:1}. 
The centre panel shows the global correlator
for an unmagnetised initial state with $m_0=0$. These data, for both $L=80$ and $L=160$, 
are much more noisy than for a larger value of $m_0$. It  is not really possible to 
extract any information from these data beyond the rough confirmation that $\Theta\approx 0.2$. 
But it appears that the short-time behaviour $C(t)\sim t^{\Theta}$ holds true irrespective of the value of $m_0$ 
such that  the practical calculation of global correlators in short-time dynamics is helped by admitting a small $m_0\neq 0$. 
Finally, in the right panel we show data for the local auto-correlator $A(t)$, 
which in principle must be studied at {\it large} and not at short times. Also, the best data are usually obtained for $m_0=0$. 
Remarkably, the data for both system sizes are practically super-imposed so that
finite-size effects do not yet arise. Using the values $\lambda=1.588(2)$ and $\mathpzc{z}=2.1667(5)$ of the critical $2D$ Ising model \cite{Okan97,Nigh00}, 
see table~\ref{tab:1}, we expect to find
$\frac{\lambda}{\mathpzc{z}}=0.7329$ which is in very good agreement with our fit $\frac{\lambda}{\mathpzc{z}}=0.74(1)$. 
In order to test the scaling relation (\ref{eq:jss}), we calculate
$\frac{\lambda}{\mathpzc{z}}=\frac{2}{2.1667} -0.190=0.733$ and find perfect agreement with our estimate, as it should be. 

We can conclude that the predictions (\ref{eq:asympt}) of short-time critical dynamics have all been very well reproduced by our numerical data on the $2D$ Ising universality class. 

\subsection{The tricritical case}

As a second example, in a distinct universality class \cite{Lawr84}, we now consider a quench onto the tricritical point $P_{\rm t}$, 
see the violet rhombus in figure~\ref{fig:2DBCphasediag}. 
We can reuse the predictions (\ref{eq:asympt}) but in the light of our experiences with $M^{(2)}(t)$ and $C(t)$ at the Ising critical point $P_c$, 
we do not display any data for $M^{(2)}(t)$ and only show data for $C(t)$ with non-vanishing $m_0\neq 0$. 

\begin{figure}[tb]
  \hspace{-1.45cm} 
    \includegraphics[width=1.18\textwidth]{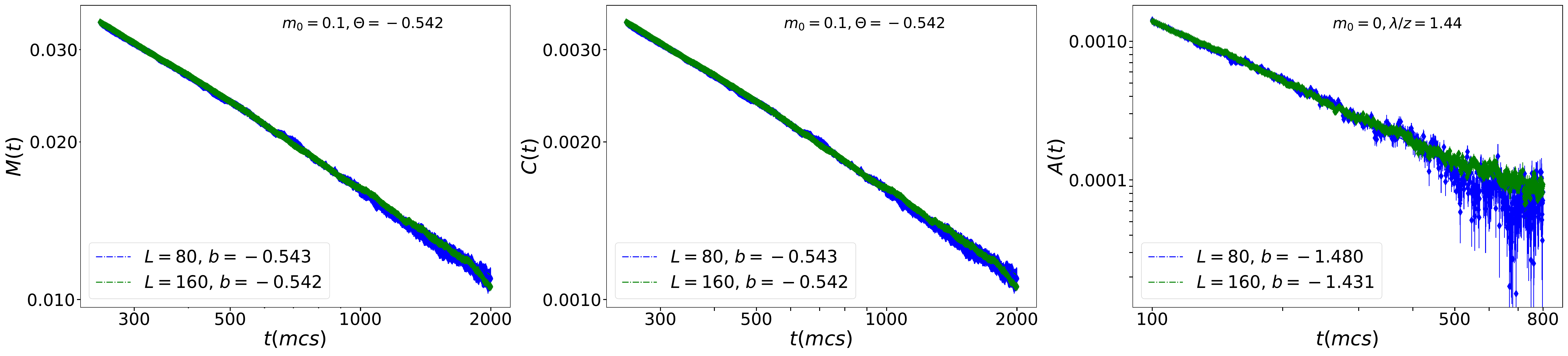}
    \caption[fiug5]{Evolution after a quench onto the tricritical point $P_{\rm t}$, with $L=80$ (blue lower curve) and $L=160$ (green upper curve) and $m_0$. 
    \underline{Left panel:} Magnetisation  $M(t)$ for $m_0=0.1$. Fit range: $t=250-2000$.
    \underline{Centre panel:} Global correlator $C(t)$, for $m_0=0.1$.  Fit range: $t=250-2000$.
    \underline{Right panel:} Local auto-correlator $A(t)$ for $m_0=0$. Fit range: $t=100-800$.
    }
    \label{fig:IsingT}
\end{figure}

In figure~\ref{fig:IsingT} we show our simulational data on the short-time tricritical behaviour. 
In the left panel, the magnetisation $M(t)$ is displayed and the central panel shows the evolution of $C(t)$, both for $m_0=0.1$. 
The $2D$ tricritical Ising model has the peculiarity that the initial slip exponent $\Theta$ is negative \cite{Jans94,Silva22a} 
and with a $2D$-value around $\Theta\approx -0.54$ \cite{Silva02,Silva13}. 
Therefore, the numerical data for $M(t)$ and $C(t)$ will actually decrease with time $t$, albeit with an exponent different from $\beta/(\nu\mathpzc{z})$. 
Since in the $2D$ tricritical equilibrium Ising model one knows from conformal invariance that $\frac{\beta}{\nu}=\frac{3}{40}$, see table~\ref{tab:1},  
this gives with $z\simeq 2.215$ \cite{Silva02} the estimate $\frac{\beta}{(\nu\mathpzc{z})}\simeq 0.034$ 
and it becomes possible to distinguish clearly a rapid short-time decay from a more slow
equilibrium-controlled decay at later times, as also mentioned in \cite{Silva02}. 
Returning to our figure, we observe in both the left and central panels a clear algebraic decay. 
For the magnetisation, the data from the different sizes are practically superimposed which means
that the scaling amplitude is essentially size-independent. 
Similarly, the middle panel for $C(t)$ shows a size-independent scaling amplitude and with the same slope. 
There is no signal for a cross-over towards the more slow late-time decay. 
For the magnetisation, a fit gives $\Theta=-0.542(5)$ and for the correlator we extract analogously the same numerical estimate $\Theta=-0.542(5)$, both
in excellent agreement with each other and with the literature \cite{Silva02,Silva13}. 
Finally, in the right panel we show data for the auto-correlator $A(t)$, with an unmagnetised initial state such that $m_0=0$. 
The data for both sizes are practically superimposed such that the scaling amplitude is size-independent, 
in agreement with theoretical expectations on dynamical finite-size effects \cite{Henk25c}. 
While the data for $L=160$ show a convincing algebraic decay, those for $L=80$ 
are considerably more noisy and their accelerated decay for large times might be viewed as a beginning of finite-size effects. 
We use that data for $L=160$ for a fit and extract $\frac{\lambda}{\mathpzc{z}}=1.43(2)$
which is consistent with the value $1.457(6)$ quoted in the literature \cite{Silva02}. 
{}From the scaling relation (\ref{eq:jss}) and the above values of $\mathpzc{z}$ and $\Theta$, we deduce
$\frac{\lambda}{\mathpzc{z}}=\frac{2}{2.215} + 0.54 = 1.44$ such that the {\sc jss} prediction is indeed confirmed. 
We are not aware of any previous explicit test of (\ref{eq:jss}) at a tricritical point.

\begin{figure}[tb] 
    \centering
    \includegraphics[width=1\linewidth]{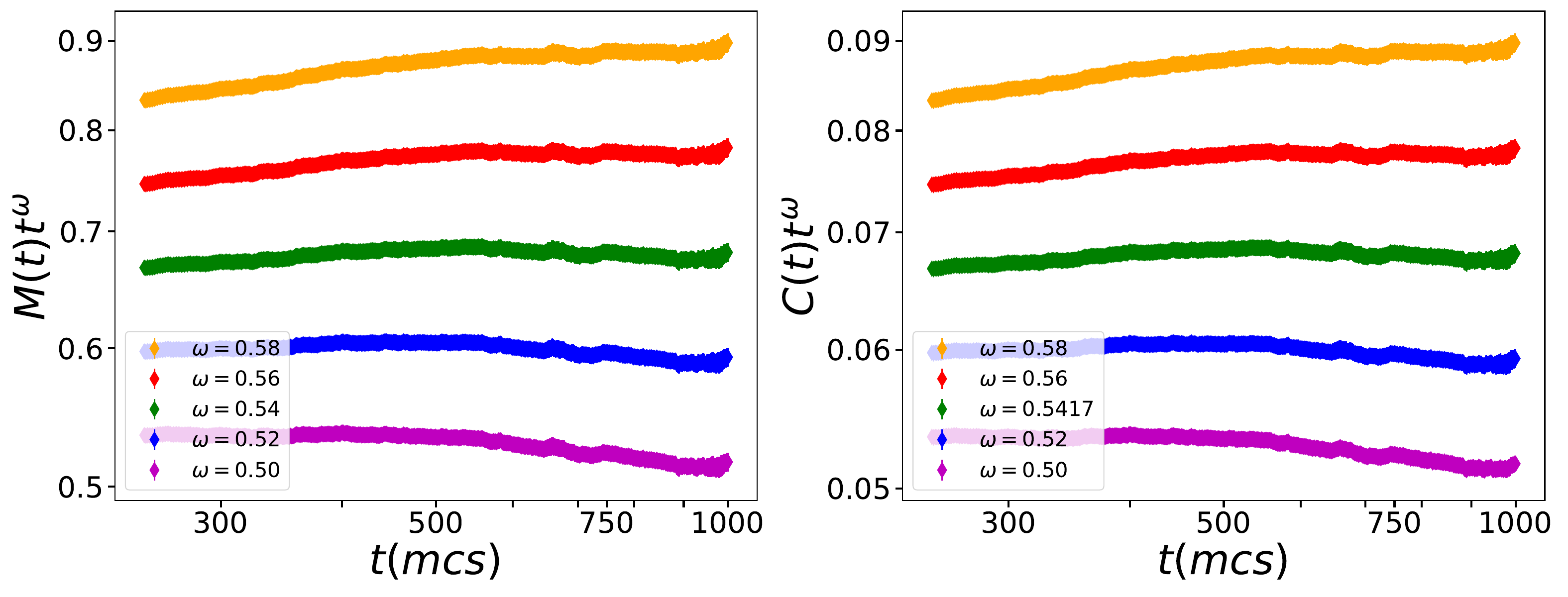}
    \caption[fig6]{Rescaled magnetisation $t^{\omega}M(t)$ (left panel) and global correlator $t^{\omega} C(t)$ (right panel) 
    in the $2D$ Blume-Capel model at the tricritical point $P_t$, for $L=160$, $m_0=0.1$, and $\omega=[0.50,0.52,0.54,0.56,0.58]$ from bottom to top.
    }
    \label{fig:IsingT-rescal}
\end{figure}

We further control these determinations by considering the rescaled magnetisation $t^{\omega} M(t)$ and the 
global correlator $t^{\omega} C(t)$ where the parameter $\omega$ is chosen freely. 
If that parameter $\omega$ is chosen such that $\omega+\Theta=0$, one expects to find a straight line, whereas the curves should increase whenever 
$\omega+\Theta>0$ and decrease for $\omega+\Theta<0$. Of course, the
conclusion reached will also depend on the time window used. In both panels of figure~\ref{fig:IsingT-rescal}, 
we find a roughly horizontal plot for the choice $\omega=0.54$, while
the is a downward curvature visible when $\omega=0.52$ is chosen and a slight upward curvature for the choice $\omega=0.56$. 
Therefore $-\Theta=\omega$ clearly falls into the interval $0.52<\Theta<0.56$ and is probably close to $0.54$. 
This confirms what we saw before from the fits. 

This concludes our study of the critical short-time dynamics in a second universality class. 
All our results are fully compatible with the established theoretical predictions (\ref{eq:asympt}) \cite{Okan97b,Zhen98,Tome98,Tome03,Alba11,Silva23} 
and existing simulational data \cite{Silva02,Silva13}. 

A final comment concerns the line of first-order transitions, see the dotted line in figure~\ref{fig:2DBCphasediag}. 
Since at a first-order transition, the correlation length remains finite, there is no reason to expect dynamical scaling to hold. 
Indeed, magnetisations and correlators all decay exponentially with time (we checked this close to the endpoint at 
$T\simeq 0$, $\Delta\simeq 2$ of the first-order line) and hence the algebraic forms of 
short-time scaling can no longer be applied. 

\section{Quenches into $T<T_c$: phase-ordering kinetics} \label{sec:4}

Having confirmed the short-time dynamics predictions for non-equilibrium dynamics, 
we can now turn to a study of what happens at short times for phase-ordering kinetics. In figure~\ref{fig:aimantation} 
we have already seen that there does exist a short-time scaling regime for quenches into the $T<T_c$ region, 
where dynamical scaling is known to hold true \cite{Bray90,Bray94a}. 
The locations of the points studied are indicated by the three blue dots below the critical line in figure~\ref{fig:2DBCphasediag}. 
Since the field-theoretical methods of {\sc jss} are not available in this case, 
we use the non-equilibrium dynamical symmetries which can be used for all $T\leq T_c$ \cite{Henk25c,Henk25e}. 
In the appendix, it is shown that this method leads to
(\ref{eq:MbasT}) for the magnetisation $M(t)$ for $m_0\neq 0$, 
but there is no known generalisation for the short-time evolutions of $M^{(2)}(t)$, $C(t)$ with $m_0\neq 0$. 
For the other observables defined in (\ref{eq:obs}) and in view of the experience gained in the
critical case, we shall use instead the known predictions \cite{Henk25c} with $m_0=0$ for the global and local correlators $C(t)$ and $A(t)$. 
Instead of (\ref{eq:asympt}), we then expect the scaling behaviours
\BEQ \label{eq:asympt0} 
M(t) \sim t^{\Theta} \;\; , \;\; M^{(2)}(t) \sim \left\{ \begin{array}{ll} t^{2\Theta}       & \mbox{\rm ~~;~ $m_0\neq 0$} \\ 
                                                                           t^{d/\mathpzc{z}} & \mbox{\rm ~~;~ $m_0=0$} 
                                                         \end{array} \right. \;\; , \;\; 
C(t) \sim t^{\Theta} \;\; , \;\; A(t) \sim t^{-\lambda/\mathpzc{z}}
\EEQ
where only the prediction for $M(t)$ is for $m_0\neq 0$ and all others are for $m_0=0$ 
unless explicitly stated otherwise.\footnote{We anticipate that the behaviour of $C(t)$ will follow (\ref{eq:asympt0}) even for $m_0\neq 0$.} 
The suggested behaviour of $M^{(2)}(t)$ for $m_0\neq 0$ is inferred from (\ref{eq:asympt}) by analogy, but in the absence of any derivation should be
considered as `{\it hypoth\`ese de travail}'. 
Since $\mathpzc{z}=2$, we have in $2D$ models that for an unmagnetised initial state with $m_0=0$ that $M^{(2)}(t)\sim t$ \cite{Janke23} which does not lead 
to an exponent estimate but rather to a consistency check of the method of short-time dynamics. 
Eq.~(\ref{eq:asympt0}) gives the short-time scaling for $M(t)$, $M^{(2)}(t)$, $C(t)$ and the long-time scaling for $A(t)$. 
We also expect that the scaling relation (\ref{eq:jss}) holds with $\mathpzc{z}=2$.

\begin{figure}[tb] 
  \hspace{-1.45cm} 
    \includegraphics[width=1.18\linewidth]{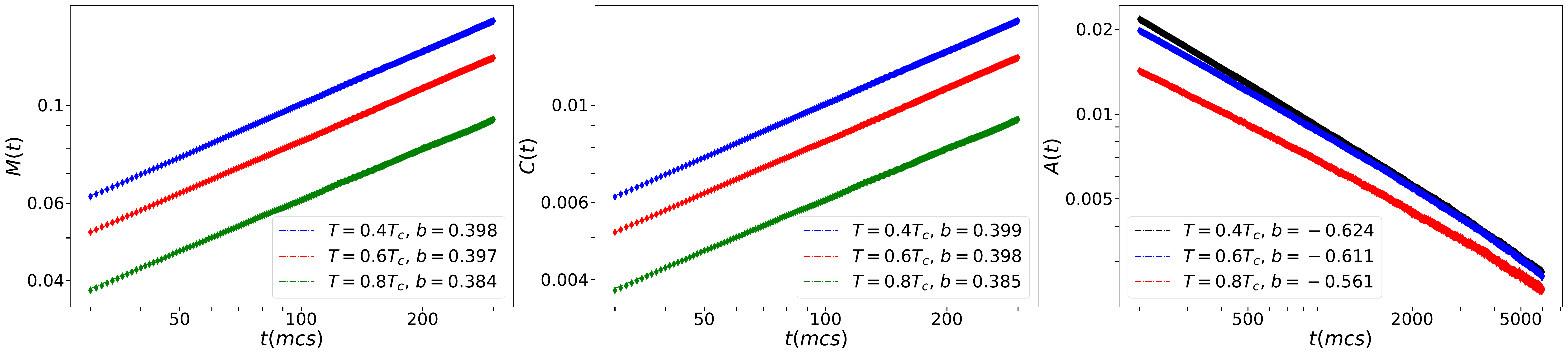}
    \caption[fig8]{Short-time dynamics in the phase-ordering of the $2D$ Blume-Capel model, for $L=160$ 
    and several values of $T=[0.4T_c, 0.6T_c,0.8 T_c]$ from top to bottom.
    \underline{Left panel:} Magnetisation $M(t)$ for $m_0=0.01$. Fit range:  $t=30-300$. 
    \underline{Centre panel:} Global correlator $C(t)$ for $m_0=0.01$.  Fit range:  $t=30-300$. 
    \underline{Right panel:} Local auto-correlator $A(t)$ for $m_0=0$. 
    Fit range:  $t=200-6000$ and with the improved ansatz $A(t) = at^{b}(1-\frac{c}{t}$) \cite{Chris20}.
    }
    \label{fig:Ising0_observablesMCA}
\end{figure}

In figure~\ref{fig:Ising0_observablesMCA} our data for three observables, namely the magnetisation $M(t)$, 
the global correlator $C(t)$ and the local auto-correlator $A(t)$ are shown, each for three different values of the temperature. 
The curves in the left and centre panels for $M(t)$ and $C(t)$ look remarkably similar. 
We find clear evidence for a power-law growth at short times (chosen sufficiently short that the cross-over to saturation is not yet visible) for these observables 
and all lines are parallel to a high degree of accuracy, which confirms the theoretical expectation that temperature should be irrelevant \cite{Bray90,Bray94a} for $T<T_c$. 
The values of the exponent $\Theta$ extracted from $M(t)$ and $C(t)$ from the left and centre panels of figure~\ref{fig:Ising0_observablesMCA} 
turn out to be almost identical, as expected from (\ref{eq:asympt0}). Averaging, we conclude
\BEQ \label{eq:Theta0}
\Theta = 0.39({1})
\EEQ
for the initial slip exponent in the phase-ordering of the $2D$ Ising universality class. It is different from the estimate obtained at criticality. 

\begin{figure}[tb] 
  \hspace{-1.45cm} 
    \includegraphics[width=1.18\linewidth]{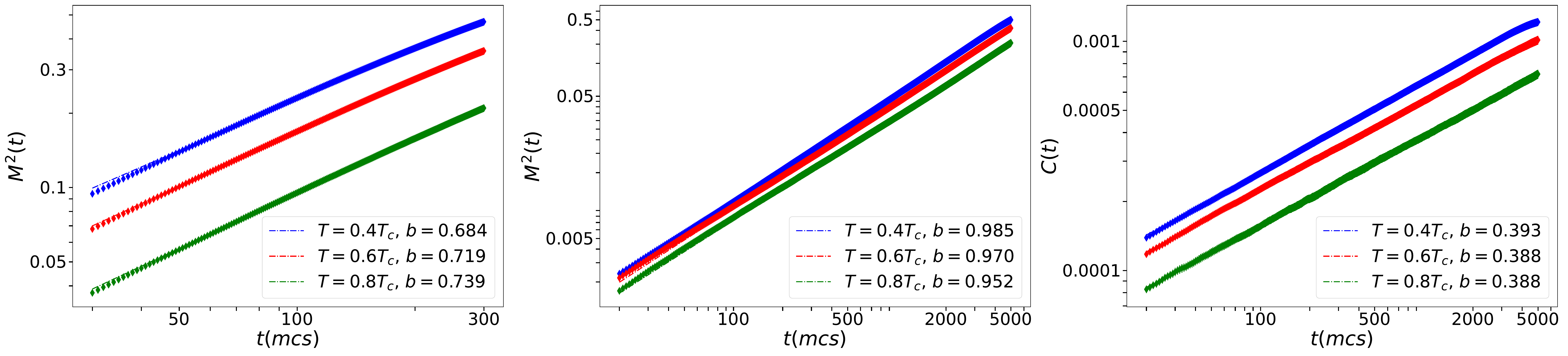}
    \caption[fig8]{Short-time dynamics in the phase-ordering of the $2D$ Blume-Capel model, 
    for $L=160$ and several values of $T=[0.4T_c, 0.6T_c,0.8 T_c]$ from top to bottom.
    \underline{Left panel:} Squared Magnetisation $M^{(2)}(t)$ for $m_0=0.05$. Fit range:  $t=30-300$. 
    \underline{Centre panel:} Squared Magnetisation $M^{(2)}(t)$ for $m_0=0$. Fit range: $t=20-5000$. 
    \underline{Right panel:} Global correlator $C(t)$ for $m_0=0$.  Fit range:  $t=20-5000$.
    }
    \label{fig:Ising0_observablesM2M2}
\end{figure}

The right panel in figure~\ref{fig:Ising0_observablesMCA} shows the evolution of the auto-correlator $A(t)$. 
In order to be able to analyse it, we must consider much larger time-scales than for the
short-time observables. Still figure~\ref{fig:Ising0_observablesMCA} shows that there does remain some considerable curvature in the data which 
impedes a simplistic and straightforward power-law fit $A(t)\sim t^{-\lambda/\mathpzc{z}}$ such that additional finite-time corrections must be taken into account. 
Since we have $\mathpzc{z}=2$ one may use the existing predictions of Schr\"odinger-invariance for the leading corrections to long-time scaling. Indeed, it can be shown that the
two-time auto-correlator takes in the limit $y\gg 1$ the form $A(ys,s)=A_{\infty} y^{-\lambda/2}\bigl( 1 - B_{\infty} y^{-1} +\ldots\bigr)$ 
and where $B_{\infty}\geq d-\lambda\geq 0$ \cite{Chris20}.\footnote{The last inequality follows from the known bounds $\frac{d}{2}\leq \lambda\leq d$ \cite{Fish88a,Yeun96a}.} 
To obtain the auto-correlator $A(t)$ with respect to the initial state, as defined in (\ref{eq:obs-A}), we shall let $s\to s_{\rm mic}$ 
go towards a microscopically small time and concentrate on the observation time $t=y s_{\rm mic}$, where
$s_{\rm mic}$ is absorbed into the constants $A_{\infty}, B_{\infty}$. 
This leads to the improved scaling $A(t) \simeq \mathscr{A} t^{-\lambda/2}\big( 1 - \mathscr{B} t^{-1}\bigr)$ with $\mathscr{A},\mathscr{B}> 0$ for large enough times. 
The results of this fit are shown in the right panel of figure~\ref{fig:Ising0_observablesMCA}. 
If the temperature $T$ is far enough from the critical point, this procedure seems to work and, 
using the data for $T=0.4 T_c$ and $T=0.6 T_c$, we extract an estimate $\frac{\lambda}{2}=0.61({1})$,
in agreement with the literature \cite{Liu91,Lorenz07a}, see also table~\ref{tab:1}. 
The data for $T=0.8 T_c$ are even more strongly curved, probably due to further notable corrections,  
and a meaningful fit does no longer seem possible. We notice that the interpretation of the 
long-time data is rendered more complicated than the one of the short-time evolution and the corrections to scaling seem to be more important in the former case. 
In order to confirm the validity of the scaling relation (\ref{eq:jss}), we use our result (\ref{eq:Theta0}) 
and calculate $\frac{\lambda}{2} = 1 - 0.39 =0.61$ which perfectly agrees with our above estimate. 

Figure~\ref{fig:Ising0_observablesM2M2} shows in the left panel $M^{(2)}(t)$ with $m_0=0.05$ and in the middle panel for $m_0=0$. In both cases, the curves are more or less 
parallel and straight, in agreement with a temperature-independent exponent $\Theta$ and a temperature-dependent scaling prefactor. 
We read off $2\Theta\simeq 0.8$ and $2/\mathpzc{z}\simeq 0.97$ which roughly support (\ref{eq:asympt0}). The right panel shows $C(t)$ for $m_0=0$ and produces once more 
$\Theta\simeq 0.39$. Remarkably, the value of $\Theta$ appears to be independent of $m_0$.

\begin{figure}[tb] 
    \centering
    \includegraphics[width=1\linewidth]{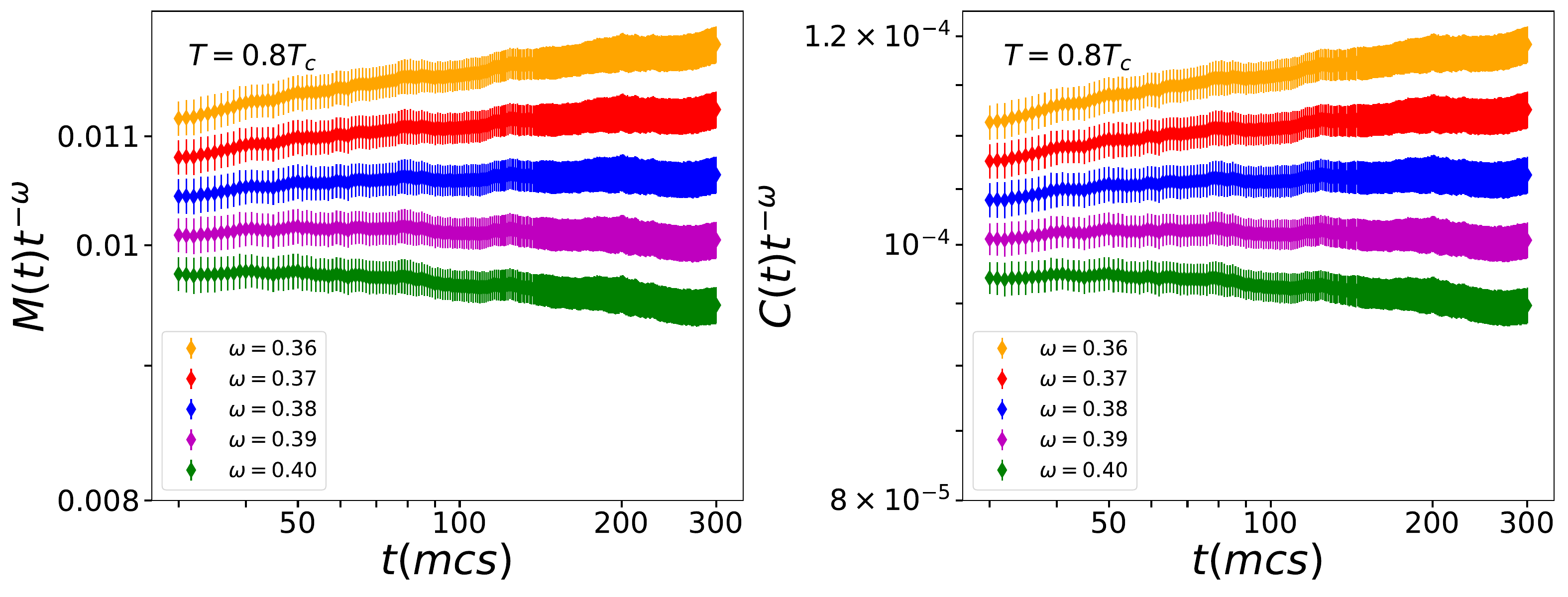}
    \caption[fig6]{Rescaled magnetisation $t^{-\omega}M(t)$ (left panel) and global correlator $t^{-\omega} C(t)$ (right panel) 
    in the $2D$ Blume-Capel model at $T=0.8T_c$, for $L=160$, $m_0=0.01$, and $\omega=[0.36,0.37,0.38,0.39,0.40]$ from top to bottom.
    }
    \label{fig:Ising08T-rescal}
\end{figure}

In figure~\ref{fig:Ising08T-rescal}, we retrace $t^{-\omega} M(t)$ and $t^{-\omega} C(t)$ 
for several values of $\omega$, over against the time $t$, for the example $T=0.8 T_c$. 
The value of $\omega$ which produces
a horizontal plot will be an estimate for the slip exponent $\Theta$. We certainly have the strict bounds $0.37<\Theta<0.40$. 
For the range of times under study, the most horizontal line
occurs for $\omega=0.38$. This means that $\Theta\approx 0.38$, which is consistent with the result (\ref{eq:Theta0}) quoted above from our fits. 

\begin{figure}[H] 
    \centering 
    \includegraphics[width=1\linewidth]{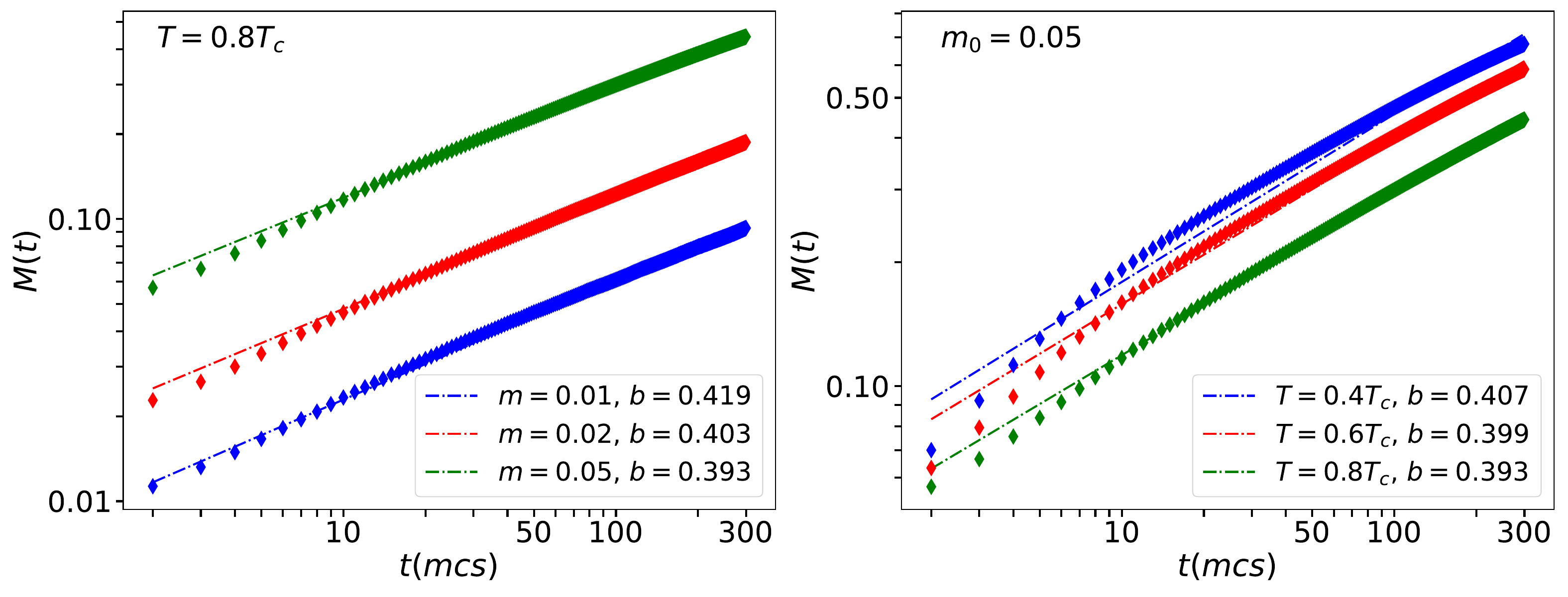}
    \caption[fig7]{Short-time magnetisation $M(t)$ in the ordered phase, for $\Delta=0$ and $L=160$.   
    \underline{Left panel:} Evolution for $T=0.8 T_c$ and $m_0=[0.01,0.02,0.05]$ from bottom to top. 
    \underline{Right panel:} Evolution for $m_0=0.05$ and $T=[0.4 T_c, 0.6 T_c, 0.8 T_c]$ from top to bottom.  
    Fit  range: $t=2-300$. 
    }
    \label{fig:Ising0-aimantation}
\end{figure}
The universality of this result in studied in figure~\ref{fig:Ising0-aimantation} 
which analyses the effect of several parameters on the short-time dynamics of $M(t)$. 
The left panel show the effect of varying $m_0$. In all cases, there is a power-law growth
with sensibly the same exponent $\Theta$, irrespective of the value of $m_0$ which confirms the expected universality. 
It is also apparent that the lattice data need to go through an initial transient regime 
before they reach time-scales where the short-time theory developed in the continuum limit can be applied. 
The effective duration of this transient regime depend on the value of $m_0$, and also on $T$ as will be discussed shortly.
In the right panel, the effect of varying the temperature $T$ is shown (these are the red dots in figure~\ref{fig:2DBCphasediag}). 
Again we find, after a brief transient regime, a power-law growth whose exponent $\Theta\approx 0.4$ is $T$-independent, 
as expected from the irrelevance of temperature for $T<T_c$ \cite{Bray90,Bray94a}, as well as independent of $m_0$. 
However, we also observe that when $T$ is decreased, the duration of the transient regime augments. 
In practise, this gives hints what values of $T$ and $m_0$  should be chosen for a reliable determination of the slip exponent. 
This confirms our conclusion (\ref{eq:Theta0}) on the value of $\Theta$
and illustrates its universality with respect to changes in $m_0$ and in $T$. 

\begin{figure}[tb]
    \centering 
    \includegraphics[width=0.6\linewidth]{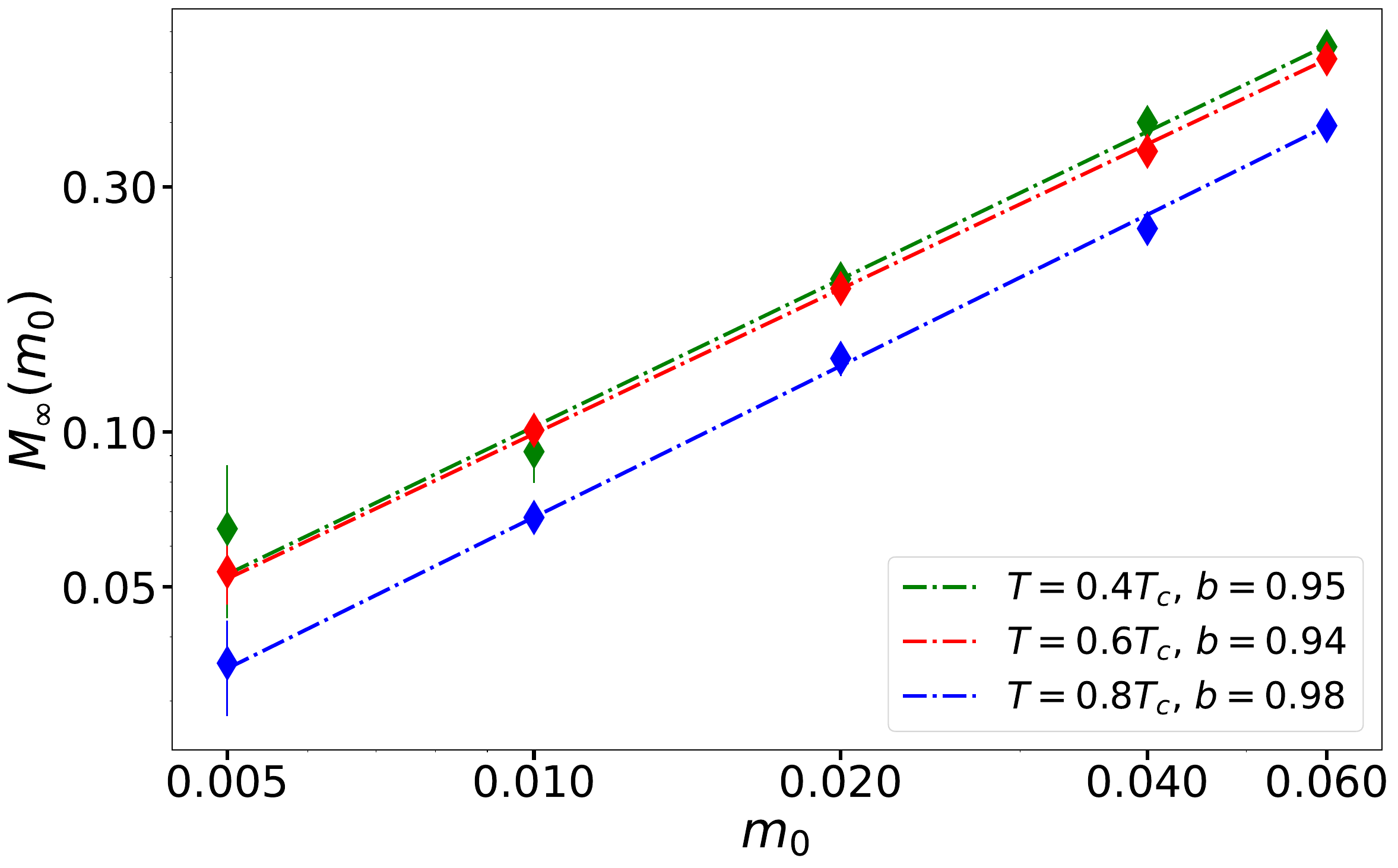}
    \caption[fig8]{Saturation magnetisation $M_\infty(m_0)$ at $t=10 000$, for $\Delta=0$ and $L=20$ 
    as a function of the initial magnetisation $m_0=[0.005-0.06]$, for temperatures $T=[0.4 T_c, 0.6 T_c, 0.8 T_c]$.}
    \label{fig:Ising0-saturation}
\end{figure}

Finally, we show in figure~\ref{fig:Ising0-saturation} the variation of the saturation magnetisation 
$M_{\infty}$ (which we estimated by calculating $M(10^4)$) with the initial magnetisation $m_0$, and for several values of the temperature. 
We find an approximate power law $M_{\infty}\sim m_0^{\mu_0}$, as expected from (\ref{eq:MbasT}). 
Herein, the exponent $\mu_0$ is roughly independent of the temperature, as can be inferred from the lines being approximately parallel. 
Averaging the results of the individual fits, we estimate $\mu_0\approx 0.96(5)$, which implies $y_0\approx 0.1$. 

In conclusion, we have presented the first example which fully satisfies (\ref{eq:asympt0}) and the scaling relation (\ref{eq:jss}) 
and thereby confirms the applicability of short-time dynamics to phase-ordering kinetics.

\section{Conclusions} \label{sec:5}

We have studied the applicability of short-time dynamics in the $2D$ Blume-Capel model, after a quench from an initially fully disordered state. 
For the well-studied non-equilibrium critical dynamics, after a quench onto (tri-)criticality, the
confirmation of the well-established scaling form (\ref{eq:JanssenM}), 
the phenomenological scaling predictions (\ref{eq:asympt}) and notably the {\sc jss} scaling relation (\ref{eq:jss}) 
was to be expected. Surprisingly, we found evi\-den\-ce 
for the existence of an analogous short-time scaling regime for phase-ordering kinetics after a quench into the ordered phase. 
Fig.~\ref{fig:aimantation} shows the respective evolutions of the time-dependent magnetisation starting from a slightly magnetised initial state 
and underscores their similarities and their differences. The scaling form (\ref{eq:MbasT}), the
phenomenological scaling predictions (\ref{eq:asympt0}) and especially the scaling relation (\ref{eq:jss}), 
linking short-time and long-time exponents, all hold true for $T<T_c$, with $\mathpzc{z}=2$. 
Table~\ref{tab:1} lists  estimates for the short-time exponent $\Theta$ and the long-time exponent $\lambda$. 
{\em The essential physical condition for the validity of these results is the existence of dynamical scaling in phase-ordering kinetics.} 

\begin{table}  
\begin{center}
\begin{tabular}{|lrr|clcc|ll|}  \hline
\multicolumn{3}{|l|}{~} & \multicolumn{4}{r|}{~} & \multicolumn{2}{r|}{~} \\[-0.4cm] 
\multicolumn{2}{|l}{model}           & ~$d$~ & ~$\Theta$~     & ~~~$\lambda$~ & ~$z$~       & ~$\beta/\nu$~ & Ref.                   & \\ \hline
\multicolumn{3}{|l|}{~} & \multicolumn{4}{r|}{~} & \multicolumn{2}{r|}{~} \\[-0.4cm] 
{\sc tdgl}         & ~$T=0$~~        & ~$1$~ & ~~~$0.199691$  & $0.600616$    & $2$         &               & \cite{Bray95}          & \\
critical Ising     & ~$T=T_c$~       & ~$2$~ & ~~~$0.190(1)$  & $1.588(2)$    & $2.1667(5)$ & $1/8$         & \multicolumn{2}{l|}{\cite{Okan97,Mend98,Nigh00}}   \\
tricritical Ising  & ~$T=T_{\rm t}$~ & ~$2$~ & $-0.542(5)$    & $3.17(4)$     & $2.215(2)$  & $3/40$        & \cite{Silva02,Silva13} & \hs{0.5}$\begin{array}{l}\mbox{\rm\scs this}\\[-0.23cm]
                                                                                                                                                                \mbox{\rm\scs work}
                                                                                                                                                    \end{array}$ \\  
Ising              & ~$T<T_c$~       & ~$2$~ & ~~~$0.39(1)$   & $1.24(2)$     & $2$         & $ $           & \cite{Liu91,Lorenz07a} & \hs{0.5}$\begin{array}{l}\mbox{\rm\scs this}\\[-0.23cm]
                                                                                                                                                                \mbox{\rm\scs work}
                                                                                                                                                    \end{array}$ \\  
Ising              & ~$T<T_c$~       & ~$3$~ &                & $1.60(2)$     & $2$         & $ $           & \cite{Henk03}          & \\ \hline 
\multicolumn{3}{|l|}{~} & \multicolumn{4}{r|}{~} & \multicolumn{2}{r|}{~} \\[-0.4cm] 
quantum Ising      & ~$h=h_c$~       & ~$1$~ & ~~~$0.3734(2)$ &               & $1$         & ~$0.125(2)$~  & \cite{Shu17}           & \\ 
quantum Ising      & ~$h=h_c$~       & ~$2$~ & ~~$0.209(4)$~  &               & $1$         & ~$0.518(1)$~  & \cite{Shu17}           & \\ \hline 
\end{tabular}\end{center}
\caption[tab1]{Values of several equilibrium and non-equilibrium exponents, in some classical and quantum systems, see also \cite{Alba11} and refs. therein. 
Notice that for $d$-dimensional critical systems $d-2+\eta=2\beta/\nu$. {\sc tdgl} is the time-dependent Ginzburg-Landau equation. 
\label{tab:1}
}
\end{table}

The conceptual basis of this work are the non-equilibrium dynamical symmetries \cite{Henk25c}. 
Is it conceivable that at least some of these might become useful beyond the context of model-A-type
dynamics explored here~? For example, in phase-separation (model-B-type dynamics with a conserved order-parameter), dynamical scaling is known to hold due to
the existence of at least two competing and equivalent equilibrium states \cite{Bray94a} but the 
order-parameter itself should remain constant \cite{Taeu14}. From long-time studies it is already known numerically 
that the long-time regime is hard to reach but the exponent $\lambda$ can be non-universal, e.g. \cite{Godr04,Muel22,Muel24}.  
On the other hand, considering models of chemical reactions without detailed balance, 
such as the noisy voter model \cite{Ligg85,Droz89,Krap92,Dorn01,Oliv03,Corb24e,Godr24}, 
directed percolation \cite{Enss04a,Ramasco04b,Baum07b,Boet18,Silva25a}
or active models \cite{Puzz22}, where a critical line is surrounded by non-critical phases with merely a single stationary state, 
the lack of dynamical scaling therein will make it impossible to apply methods of short-time dynamics. 

Quantum dynamics is another promising field for further applications. For example, 
in closed quantum systems one may study imaginary-time dynamics through a Schr\"odinger equation
$\hbar\partial_{\tau} \ket{\Psi(\tau)} = H \ket{\Psi(\tau)}$ where $H$ 
is a quantum hamiltonian and $\ket{\Psi(\tau)}$ a quantum mechanical state in imaginary time $\tau$. 
We include in table~\ref{tab:1} results for the critical initial slip exponent $\Theta$ obtained
for the quantum Ising model in a transverse field $h$ at the quantum critical point $h_c$ (and at vanishing temperature $T=0$) \cite{Shu17}. 
The equilibrium critical behaviour in $d_{\rm qu}$ quantum dimensions is related to the one of a classical equilibrium system in 
$d_{\rm cl}=d_{\rm qu}+\mathpzc{z}_{\rm qu}$ classical dimensions \cite{Sach11}. 
The non-equilibrium values of the slip exponent, however, of the quantum and classical systems are different and might appear unrelated. 
Remarkably, from the data in table~\ref{tab:1} the following conjecture can be formulated
\BEQ \label{conj}
\lambda_{\rm cl} \stackrel{?}{=} 2 - (\mathpzc{z}_{\rm qu}+1) \Theta_{\rm qu}
\EEQ
If confirmed, this would relate the long-time classical exponent $\lambda_{\rm cl}$ in the ordered phase to the short-time quantum critical exponent 
$\Theta_{\rm qu}$ and the quantum dynamical exponent $\mathpzc{z}_{\rm qu}$.
The proposed relationship is bourne out by the pair 
\BD 
\begin{array}{l} \mbox{\rm low-temperature classical Ising model} \\ 
                 \mbox{\rm  in $d+1$ dimensions}
\end{array}
~~~\Longleftrightarrow~~~ 
\begin{array}{l} \mbox{\rm critical quantum Ising model} \\
                 \mbox{\rm in $d$ dimensions, at $T=0$}
\end{array}
\ED 
for $d=1$ and $d=2$. 
For the time being, this is just a numerical observation.\footnote{If (\ref{conj}) were true, 
the estimate $\Theta_{\rm qu}=0.3734(2)$ \cite{Shu17} would give the best existing confirmation of the Fisher-Huse conjecture $\lambda_{\rm cl}=\frac{5}{4}$ \cite{Fish88a}.} 
Further work will be needed to either confirm or invalidate this conjecture. 
Possible extensions to open quantum systems, e.g. \cite{Ding25,Zahr25}, would be of great interest.

\noindent
{\bf Acknowledgements:}  
{\sc lm} was supported by Coventry University, the L4 collaboration Leipzig-Lorraine-Lviv-Coventry and the UFA 
(Universit\'e Franco-Allemande) through the Coll\`ege Doctoral franco-allemand. {\sc mh} was supported by the french ANR-PRME UNIOPEN (ANR-22-CE30-0004-01).

\newpage 

\appsektion{Theoretical basis for short-time dynamics}
The derivation of the time-dependent magnetisation $M(t)$, notably (\ref{eq:MbasT}) together with (\ref{eq:jss}), 
requires to analyse the scaling of the two-time response function $R$ 
of the order-parameter $\phi$ with respect to a perturbation by the conjugate magnetic field $h$. 
Assuming spatial translation-invariance, the response function is
\BEQ
R(t,s;\vek{r}) = \left. \frac{\delta \bigl\langle \phi(t,\vek{r})\bigr\rangle}{\delta h(s;\vek{0})}\right|_{h=0} 
= \left\langle \phi(t,\vek{r})\wit{\phi}(s,\vek{0}) \right\rangle
\EEQ
where $s$ is the waiting time, $t>s$ the observation time and we recall from Janssen-de Dominicis theory (we restrict exclusively to model-A dynamics) that 
$R=\bigl\langle \phi \wit{\phi}\bigr\rangle$ can be thought of as a correlator of the order-parameter $\phi$ 
with the associated response scaling operator $\wit{\phi}$ \cite{Jans76,Domi76}.  

Physical ageing occurs far from equilibrium and is by definition not time-trans\-la\-tion-in\-va\-ri\-ant. 
The generators of the Lie algebra of dynamical symmetries are mapped onto a non-equilibrium representation and
without standard time-translation-invariance, by the following 

\noindent
{\bf Postulate:} \cite{Henk25c} {\it The Lie algebra generator $X_n^{\rm equi}$ of a time-space symmetry of an equilibrium system becomes a symmetry 
out-of-equilibrium by the change of representation}
\BEQ \label{gl:hyp}
X_n^{\rm equi} \mapsto X_n = e^{\xi \ln t} X_n^{\rm equi} e^{-\xi \ln t}
\EEQ
{\it The parameter $\xi$ characterises the scaling operator $\phi$ on which $X_n$ acts.} 

For our purposes, it it sufficient to restrict our attention to dilatations and time-translations. 
For the dilatation generator $X_0^{\rm equi}=-t\partial_t-\delta$ the prescription (\ref{gl:hyp}) merely leads to a modified scaling dimension 
$\delta_{\rm eff} = \delta-\xi$, since 
\begin{subequations} \label{gl:Xgen}
\BEQ \label{gl:X0gen} 
X_0^{\rm equi} \mapsto X_0 = -t\partial_t - \frac{1}{\mathpzc{z}}r\partial_r - \bigl(\delta - \xi\bigr)
\EEQ
However, the time-translation generator $X_{-1}^{\rm equi}=-\partial_t$ now becomes
\BEQ \label{gl:X-1gen}
X_{-1}^{\rm equi} \mapsto X_{-1} = -\partial_t + \frac{\xi}{t}
\EEQ
\end{subequations}
In this new non-equilibrium representation, the `physical' scaling operators become 
$\phi(t) = t^{\xi} \phi^{\rm equi}(t) = e^{\xi \ln t}\phi^{\rm equi}(t)$ and are characterised by a pair $(\delta,\xi)$ of scaling dimensions. 
Analogously, a response scaling operator $\wit{\phi}$ is characterised by a pair $(\wit{\delta},\wit{\xi})$. 
The Bargman super\-se\-lec\-tion rules \cite{Barg54} impose that $\delta=\wit{\delta}$, but $\xi$ and $\wit{\xi}$ remain independent \cite{Henk25c,Henk25e}.  

The applicability of this postulate to ageing, for all $T\leq T_c$, was studied for two-time auto-responses and auto-correlators assumed covariant under $X_{-1,0}$ in \cite{Henk25,Henk25c}. 
The inclusion of single-time correlators is achieved by restricting the requirement of covariance to the response functions only, but then the Lie algebra can be extended to the 
full Schr\"odinger algebra \cite{Henk25d,Henk25e,Henk25f}. All generic properties of ageing can be deduced from these symmetry requirements. 
Specifically, for phase-ordering kinetics one finds, when the initial correlations are spatially short-ranged, from the response function $R=\bigl\langle \phi\wit{\phi}\bigr\rangle$, 
that the autocorrelation or auto-response exponent $\frac{\lambda}{2}=\delta = \xi = -\wit{\xi}$ \cite{Henk25c}.

We are interested in the behaviour of the time-dependent magnetisation $M(t)$, 
starting from an initial slightly magnetised state with initial magnetisation $m_0\ll 1$ but otherwise uncorrelated. At the critical point $T=T_c$, this was studied by
Janssen, Schaub and Schmittmann ({\sc jss}) long ago \cite{Jans89}, and re-derived recently from the dynamical symmetries (\ref{gl:Xgen}) \cite{Henk25,Henk25c}. 
Here, we present the extension of the argument to phase-ordering kinetics when $T<T_c$. 

The \underline{first step} consists of relating $M(t)$ to a (global) response function, leading to (\ref{gl:m0R}). 
This follows from Janssen-de Dominicis non-equilibrium field-theory \cite{Domi76,Jans76,Jans92,Taeu14},
which in principle finds averages of an observable $\mathscr{A}$ as a functional integral 
\BEQ \label{dynft}
\bigl\langle \mathscr{A}\bigr\rangle = \int \mathscr{D}\phi\mathscr{D}\wit{\phi}\; \mathscr{A}[\phi]\, e^{-{\cal J}[\phi,\wit{\phi}]}
\EEQ
where the action ${\cal J}[\phi,\wit{\phi}]={\cal J}_0[\phi,\wit{\phi}]+{\cal J}_{\rm ini}[\wit{\phi}]$ depends on both the order-parameter $\phi$ 
and the conjugate response operator $\wit{\phi}$ of which the latter describes the coupling of the system's evolution with the initial correlations. 
The action is decomposed into a deterministic term and an initial term
(the explicit form (\ref{dynact}) is for non-conversed model-A dynamics)  
\begin{subequations} \label{dynact}
\begin{align}
{\cal J}_0[\phi,\wit{\phi}] &= \int \!\D t\D\vek{r}\: \left( \wit{\phi} \left( \partial_t - \Delta_{\vek{r}} - V'[\phi]\right)\phi  \right) \\
{\cal J}_{\rm ini}[\wit{\phi}] &= \int \!\D\vek{r}\: \left( \frac{1}{2\tau_0} \wit{\phi}^2(0,\vek{r}) - m_0 \wit{\phi}(0,\vek{r}) \right) \label{dynact-ini}
\end{align}
\end{subequations}
(with the usual re-scalings) which describes an initially magnetised state of average magnetisation $m_0$ together with gaussian
fluctuations of width $\tau_0$. Since the temperature is known to be irrelevant in phase-ordering kinetics \cite{Bray90,Bray94a} it is not included here. 
The non-trivial short-time relationship $\phi(0,\vek{r})=\frac{1}{\tau_0}\wit{\phi}(0,\vek{r})$ \cite{Jans92} is implicit 
and `initial time' refers to a time-scale $\tau_{\rm mic}$ at the beginning of the scaling regime. 
We consider {\em deterministic averages} $\bigl\langle \cdot \bigr\rangle_0$ defined as
\BEQ
\bigl\langle \mathscr{A}\bigr\rangle_0 = \int \mathscr{D}\phi\mathscr{D}\wit{\phi}\; \mathscr{A}[\phi]\, e^{-{\cal J}_0[\phi,\wit{\phi}]}
\EEQ
Either from causality considerations \cite{Jans89,Jans92,Taeu14} or from combined Galilei- and spatial trans\-la\-tion-in\-va\-ri\-an\-ce of ${\cal J}_0$ \cite{Pico04} 
one obtains the Bargman su\-per\-se\-lec\-tion rules \cite{Barg54}
\BEQ \label{Bargman}
\left\langle \overbrace{~\phi \cdots \phi~}^{\mbox{\rm ~~$n$ times~~}} 
             \overbrace{ ~\wit{\phi} \cdots \wit{\phi}~}^{\mbox{\rm ~~$m$ times~~}}\right\rangle_0 \sim \delta_{n,m}
\EEQ
for the deterministic averages. Only observables built from an equal number of order-parameters $\phi$ 
and conjugate response operators $\wit{\phi}$ can have non-vanishing deterministic averages. 

Finding $M(t)$ from (\ref{dynft}), we include the initial action ${\cal J}_{\rm ini}[\wit{\phi}]$ into the deterministic average. 
Expanding then the exponential to all orders in $m_0$, the
Bargman rule (\ref{Bargman}) implies that a single term survives such that the time-dependent magnetisation can be obtained as \cite{Pico04,Henk10}
\BEQ \label{gl:m0R}
M(t) = \bigl\langle \phi(t,\vek{0})\bigr\rangle 
= m_0 \int_{\mathbb{R}^d} \!\!\D\vek{r}\: \bigl\langle \phi(t,\vek{0}) \wit{\phi}(0,\vek{r}) \bigr\rangle_0 = m_0 \int_{\mathbb{R}^d} \!\!\D\vek{r}\: R(t,0;\vek{r}) 
= m_0 \wht{R}(t,0;\vek{0})
\EEQ
By definition, this is a global response function, defined by the Fourier transform
\BEQ \label{Fourier} 
\wht{R}(t,s;\vek{q}) = \int_{\mathbb{R}^d} \!\D\vek{r}\; e^{-\II \vek{q}\cdot\vek{r}} R(t,s;\vek{r})
\EEQ
The formally vanishing waiting time $s=0$ in (\ref{gl:m0R}) should be physically interpreted as a microscopically small
time $s_{\rm mic}\ll t$ at the beginning of the scaling regime. 
Non-linearities in ${\cal J}_0[\phi,\wit{\phi}]$ are irrelevant for large enough times when $\lambda>1$ \cite{Henk25c,Henk25e}. 

In the \underline{second step}, we now derive $M(t)$ from (\ref{gl:m0R}). 
This will give eq.~(\ref{eq:MbasT}), along with (\ref{eq:jss}), in the main text. 
The initial magnetisation $m_0$ has a scaling dimension related to $y_0$. 
The co-variance conditions of the global response function $\wht{R} = \wht{R}(t,s;\vek{0};m_0)$ then read
\begin{subequations} \label{gl:m0}
\begin{align}
X_{-1} \wht{R} &= \left( -\partial_t - \partial_s + \frac{\xi}{t} - \frac{\xi}{s} \right) \wht{R} =  0 \label{gl:m0-1} \\
X_{0}  \wht{R} &= \left( -t\partial_t - s\partial_s + \frac{1}{y_0}  m_0\partial_{m_0} -2\delta +\frac{d}{2} \right) \wht{R} = 0 
\label{gl:m00}
\end{align}
\end{subequations}
and we recall that $\mathpzc{z}=2$ \cite{Bray94b} and $\delta=\xi=-\wit{\xi}$ \cite{Henk25c} in phase-ordering kinetics. 
Solving the differential equations (\ref{gl:m0}) is straightforward \cite{Kamk79} and gives for $t/s\gg 1$ the asymptotics
\BEQ \label{A11}
\wht{R}(t,s;\vek{0};m_0) = 
s^{-2\delta+d/2} \left( \frac{t}{s}\right)^{-\delta+d/2} \mathcal{F}\left(m_0\, t^{1/y_0}\right)
\EEQ
with a scaling function $\mathcal{F}(u)$ of the single argument $u=m_0\, t^{1/y_0}$. 
The global response (\ref{gl:m0R}) is effectively with respect to an `initial' perturbation. 
To take this into account, we send $s\to s_{\rm mic}$ to a microscopically time-scale and then can absorb it into the scaling function. 
Combination of (\ref{A11}) with (\ref{gl:m0R}) gives the sought time-dependent magnetisation 
(recall $\delta=\frac{\lambda}{2}$ \cite{Henk25c,Henk25e}) 
\BEQ \label{36}
M(t) = m_0\: t^{\Theta} \mathscr{F}_M\left( m_0^{y_0}\: t\right) \;\; , \;\;
\Theta = \frac{d}{2}-\delta = \frac{d-\lambda}{2}
\EEQ
which is (\ref{eq:MbasT}) in the text. For short times, the magnetisation scales as $M(t)\sim t^{\Theta}$ 
and, with $\mathscr{F}_M(0)=\mbox{\rm cste.}$ (required  since $M(t)$ should be linear in $m_0$ for $t$ small) 
we recognise the analogue of the initial slip identified at criticality by {\sc jss} \cite{Jans89} and illustrated in figure~\ref{fig:aimantation} in the text. 
We also see that the slip exponent $\Theta$ in (\ref{36}) satisfies the same relation 
(\ref{eq:jss}) \cite{Henk25c,Henk25e} with the auto-correlation exponent $\lambda$ as at the critical point, if
we use that $\mathpzc{z}=2$ for $T<T_c$ in phase-ordering kinetics \cite{Bray94b}. 
We have thereby generalised (\ref{eq:JanssenM},\ref{eq:jss}) to all temperatures $T<T_c$. The scaling function $\mathscr{F}_M$ interpolates between
the two regimes of non-equilibrium critical scaling. The $t$-independent saturation observed in figure~\ref{fig:aimantation} (right panel) 
for large times is captured by admitting $\mathscr{F}_M(u)\sim u^{-\Theta}$ for $u\gg 1$. 
In contrast to the critical case, for $t\to\infty$ we now have a plateau value $M_{\infty}\sim m_0^{\mu_0}$ 
with  $\mu_0={1-\Theta y_0}$. Only this very last step of matching the scaling function offers any essential difference of the argument with respect to the critical case. 

This derivation is restricted to the magnetisation $M(t)$ and cannot be easily extended neither to higher moments $M^{(n)}(t)$, nor correlators, 
when $m_0\neq 0$.\footnote{This would probably require to study four-point responses, see \cite{Henk25e}.} 

This is quite analogous to the initial slip at criticality \cite{Jans89}, originally only derived for model-A dynamics, 
but later extended to model-C-dynamics as well \cite{Oerd93,Cala03}, 
or to anti-ferromagnets \cite{Nand19,Nand20}. However, Schr\"odinger-covariance of the response function  is not 
sufficiently strong to derive the form of $\mathscr{F}_M(u)$ \cite{Henk25e}. 
We leave this as an open problem. In searches for such a larger symmetry known exact results, 
e.g. in the spherical model \cite{Dieh97,Dutt08},  might become useful.


\end{document}